# Towards anomaly detection in smart grids by combining Complex Events Processing and SNMP objects

Massimiliano Leone Itria
*ResilTech S.r.l.*
Pontedera (PI), Italy
massimiliano.itria@resiltech.com

Enrico Schiavone
*ResilTech S.r.l.*
Pontedera (PI), Italy
enrico.schiavone@resiltech.com

Nicola Nostro
*ResilTech S.r.l.*
Pontedera (PI), Italy
nicola.nostro@resiltech.com

*Abstract*— This paper describes the architecture and the fundamental methodology of an anomaly detector, which by continuously monitoring Simple Network Management Protocol data and by processing it as complex-events, is able to timely recognize patterns of faults and relevant cyber-attacks. This solution has been applied in the context of smart grids, and in particular as part of a security and resilience component of the Information and Communication Technologies (ICT) Gateway, a middleware-based architecture that correlates and fuses measurement data from different sources (e.g., Inverters, Smart Meters) to provide control coordination and to enable grid observability applications. The detector has been evaluated through experiments, where we selected some representative anomalies that can occur on the ICT side of the energy distribution infrastructure: non-malicious faults (indicated by patterns in the system resources usage), as well as effects of typical cyber-attacks directed to the smart grid infrastructure. The results show that the detection is promisingly fast and efficient.

## I. Introduction

A smart grid is an electricity supply network based on Information and Communication Technologies (ICT) and digital information, which connects suppliers and prosumers allowing monitoring, analysis and control of many aspects of production and distribution of electricity [1]. Smart grids represent one of the systems, among the emerging ones, which show a certain degree of criticality where the effective and prompt detection of both malfunctions and intentional cyber-attack attempts is of critical importance [2]. Nowadays, in fact, cyber security in smart grid systems is becoming a major concern, since there is a wide list of cyber-attacks that can be conducted against them [3], and some of them have been already demonstrated to be actually feasible [4]. Solutions have been proposed which promise to mitigate the risk of cyber-attacks [4][5][6] and faults [7], but we believe that there is still space for new methods capable of timely detecting the occurrence of undesired events caused by intentional attacks or faults in the system, and we cannot rely on existing mitigations forever due to a continuously evolving landscape in cyber security and technology (e.g., emerging threats).

In this work, we present a general approach for fault and attack detection in ICT networks, specifically designed for being applicable to smart grid systems, which exploits anomaly detection [8] to find irregular values in monitored Simple Network Management Protocol (SNMP) [9] data, and combines it with Complex Event Processing (CEP) [10] for correlating those anomalous values to known cyber-attacks and faults.

The paper is structured as follows. Section II reports background concepts of SNMP and CEP. Section III gives an overview of the context in which the solution has been designed. Section IV describes the proposed approach. Section V reports the experiments executed and their results. Finally, Section 0 draws conclusion and outlines future work.

## II. Background: SNMP and CEP

### A. The SNMP

The SNMP Protocol [9] is an Internet Standard protocol for the management and monitoring of network and computer equipment.

It is supported by many typical network devices (e.g., routers, hubs, bridges, switches, servers, modem and racks). SNMP standard includes an application layer protocol based on a manager entity (i.e. management system) that is responsible to communicate with the agents implemented in the network devices. The management system is able to get relevant information, called SNMP objects [9] (e.g., resource usage, running processes, number of packets sent/received), from the network devices that support the protocol.

The use of SNMP data has been proposed as means to detect Denial of Service (DoS) attacks in computer network through statistical algorithms [11]. The advantage of using SNMP data is that this information is easily available as the protocol is enabled in the involved network equipment to be monitored.

### B. The Complex Event Processing

CEP [10] is a set of concepts and techniques that consists in the real-time or nearly real-time processing of events generated by the combination of data from multiple sources and aggregated in complex-events representing something occurred in a system. Common event processing operations include reading, creating, transforming and deleting events. CEP is the means that allows to: i) filter and extract relevant events from several data streams belonging to event producers; ii) identify relationships; iii) correlate such events, and; iv) aggregate the information in complex-events. Therefore, the most important feature of CEP is its ability to analyze many events and possibly recognize patterns of spotted complex events very quickly [12].

III. THE APPLICATION CONTEXT: THE ICT GATEWAY

This section introduces the overall architecture and the application context for the use of the proposed solution. The ICT Gateway (ICT GW) is a software system that acts as a mediator between data sources (i.e., headends) and grid actuators, providing services for smart energy distribution and ensuring security and resilience of the system.

The middleware-based architecture for data fusion, referred as ICT GW [13], is Fig. 1. The ICT GW leverages smart grid topology information jointly with data provided by measurement devices deployed in the field, for achieving grid observability. It uses existing communication technologies and off-the-shelf computing hardware to obtain novel control coordination from measurement functionality. The measurements from different headends, such as Inverters, Smart Meters, via an AMI (Advanced Metering Infrastructure), and Electrical Measurement in general, are correlated with data originating in existing Distribution System Operator (DSO) subsystems, such as the grid topology, enabling and developing novel grid observability applications (e.g., for outage detection and diagnosis, grid operation efficiency, and voltage quality) [14][15]. Then, the gained observability can be exploited to control and coordinate, thanks to the actuation capabilities offered by the inverters and together with existing DSO actuation, to enhance voltage quality and minimize losses in the Low Voltage (LV) grid [16].

The ICT GW is thus a middleware layer that provides a unique framework for domain-oriented applications by abstracting specific details regarding subsystem interface into a common and harmonized data model. One of the roles of the ICT GW is also to take care of reliability and security issues through the Security & Resilience component where the anomaly and attack detector proposed in this work takes place.

IV. THE PROPOSED DETECTION APPROACH

The approach adopted in this work involves the runtime analysis and correlation of SNMP objects through a correlator engine, allowing the DSO, or the ICT network administrator, to implement detection patterns and enable them for monitoring the network in near real-time.

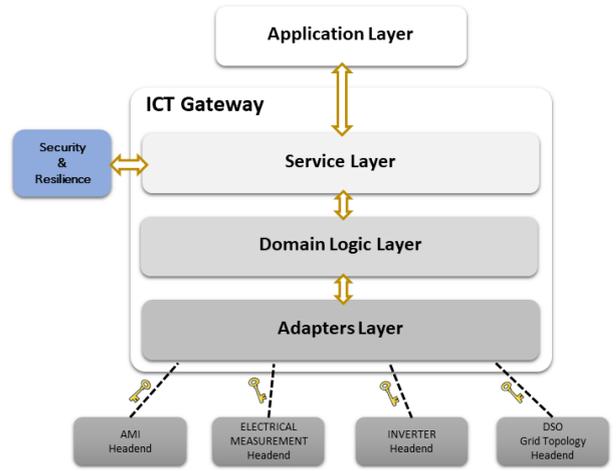

Fig. 1. Architecture of the ICT Gateway [13]

The idea is to monitor the trends of a set of selected SNMP objects and to correlate them based on detection patterns, thus allowing to recognize anomalies or faults occurring in the network.

Leveraging the CEP, once either a fault or an attack pattern is recognized a warning is immediately raised for activating the recovery system in charge of actuating the proper reaction to solve or mitigate the emergent criticality.

*A. Architecture of the Anomaly Detector*

The idea behind the anomaly detector introduced in this work starts from continuously monitoring SNMP objects of the ICT Network (i.e., the ICT GW in our context). The information obtained through SNMP objects is compared with fault and attack detection patterns by means of CEP technology. Patterns can be defined by studying the effect of the attacks on the SNMP objects trend and identifying relationships between measures and thresholds. It is possible to define several patterns for the same type of attack.
In this work, attacks have been reproduced in lab and an injection activity has been conducted in order to analyze the variation in SNMP objects and deduce representative, general, patterns.

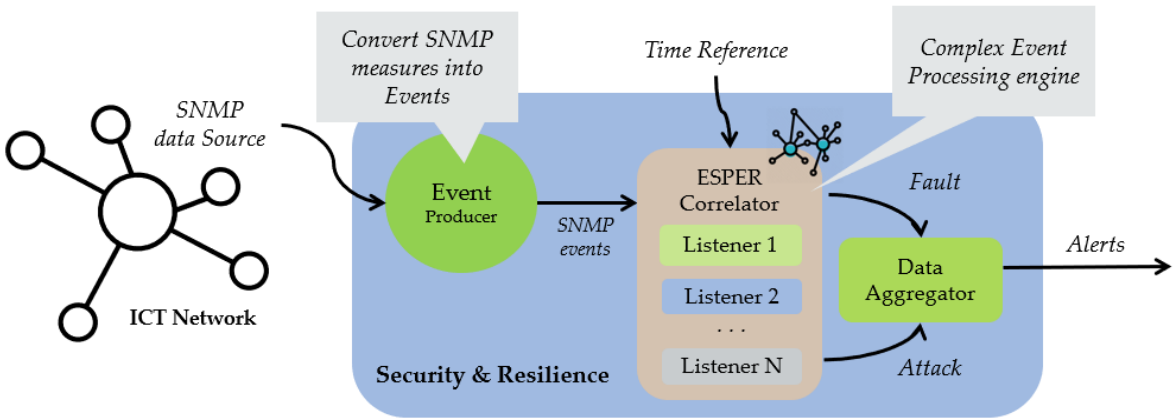

Fig. 2. The Security and Resilience (S&R) component architecture

Finally, after the deployment of the anomaly detector, a new set of attacks has been injected in order to verify the effectiveness in detection.

The anomaly detection is implemented as part of the Security and Resilience (S&R) functionalities. The S&R is a component that performs both attack and fault detection for the ICT GW. As depicted in Fig. 2., the S&R takes in input SNMP objects, converts them into events, and forwards the events to the CEP engine, namely the Esper Correlator [17].

The Esper Correlator works, to a certain extent, as a reversed database: instead of storing data and running queries on data, it allows applications to store queries and run them on the data streams. The Correlator receives SNMP objects from the ICT GW, applies queries to the data stream and raises alerts when conditions occur that match the queries. Queries are expressed through the Event Processing Language (EPL) [17], a domain specific language designed for similarity with SQL (Structured Query Language), with the difference that EPL uses stream views rather than tables, and thus allows for the processing of several streams within a single query. Detection patterns are implemented as Esper queries activated on the SNMP objects stream.

In the Correlator, each query is activated in a Listener, i.e. a software module in charge of applying the query on the input data. Once a query matches, the listener triggers the Data Aggregator that provides in output to the Operator the related alert describing the anomaly originating it.

## V. EXPERIMENTAL EVALUATION AND RESULTS

Intentional malicious actions, as well as accidental faults, for example caused by HW failures, SW bugs, inadequate maintencance, etc., occurring in the ICT side of the electrical distribution infrastructure, could manifest themselves through deviations from the expected behavior.

Therefore, monitoring system resources (e.g., RAM (Random Access Memory) usage, storage requests, CPU (Central Processing Unit) workload) can help detecting dangerous, non-malicious, situations before malfunctions occur or at worst before the service disruption. On the other hand, anomalies in system resources (e.g., CPU and/or RAM overload) can also indicate intrusions which exploited a vulnerability. In both cases, patterns for resources monitoring allow the detector to detect when a system resource is overloaded and can be a good indicator of a dangerous situation to be taken into consideration.

Together with system resources, anomalies in the network activity may be as well indicative of deviations from nominal behaviors.

Among many types of attacks affecting smart grids ICT infrastructure [3][4][18], we focused on DoS attacks, which are one of the main threats to smart grids communications [19][20]. However, the proposed approach is suitable for identifying and predicting several type of attacks. Indeed, by means of specific SNMP objects, it is possible to monitor network access points, resource usage and active processes (e.g., memory, storage, kernel processes), thus allowing to detect and prevent network intrusion attempts, malicious code corruption or malicious code activation, which could allow the attacker to gain control of the system.

For demonstration purposes only, the TCP (Transmission Control Protocol) Reset attack and the ICMP (Internet Control Message Protocol) Flooding attack have been selected as representatives of a large number of common DoS attacks [21][22] and used for the experimental evaluations presented in this paper.

The TCP Reset attack has the purpose of breaking an existing connection between two victim hosts and spoofing a single TCP RST (details are given section B) packet may be sufficient to achieve the goal [22]. In the case of an ICMP Flooding, an attacker sends a large number of spoofed ICMP packets from a very large set of source IP. When a server is flooded with massive amounts of spoofed ICMP packets, its resources are exhausted in trying to process these requests. This overload reboots the server or has a huge impact on its performance [21].

The experiments have been performed using a physical machine on which both the ICT GW and the S&R component run. On the other hand, a virtual machine simulated the attacker's endpoint from where the attacks have been launched.

In our experiments, the detector has been implemented and trained by injecting two representative attacks and observing patterns and changes in the SNMP objects while the attacks where occurring. The results of this preliminary campaign have been used to define the detection queries to be included in the correlator.

Afterwards, attacks different from the ones injected during the design of the detector, but belonging to the same categories, where injected to test the detection effectiveness. It is worth noting that the injection campaign aimed at validating the solution has been performed by means of different tools and configurations.

However, different or additional patterns and objects can be adopted to improve the detection capabilities. In fact, the operator can adjust thresholds according to the evolution of known attacks or introduce completely new patterns for start detecting other ones.

### A. System Resources Anomaly Detection

The tests conducted to detect resources overload were performed considering three different SNMP objects.

- *hrProcessorLoad* for alerting when the CPU usage is over a predefined threshold that can cause services slowing down.

- *hrStorageUsed* for detecting RAM overload by setting a threshold on the maximum amount of usable memory that allows the system to operate without slowing down.

- *hrSystemProcesses* for tracing the number of tasks activated simultaneously on the machine.

These three objects are sufficient to detect the faults referred in this study, but this does not exclude that different SNMP objects can be combined to detect the same type of faults.

In our tests, based on heuristic approach, we set *hrProcessorLoad* threshold to 90%, considered a balanced amount of CPU usage that allows the ICT GW to operate efficiently without slowing down. Similar tests have been executed on hrStorageUsed for determining the RAM limits.

In normal conditions, several processes are activated on the ICT GW, either tasks related to the services or tasks related to the operating system running on the machine where the ICT GW is placed. Therefore, the number of tasks varies over time mainly because the operating system schedules automatically processes that do not depend on the services. However, we experienced that, under normal conditions, this number does not exceed 40 simultaneously active tasks. In this experiment, an alert has been raised by the S&R when *hrSystemProcesses* was above the 40 tasks threshold.

These alerts represent anomalies occurred in the system and they cannot be associated a priori with specific accidental faults or malicious intentional actions. However, such anomalies are symptoms that should not be underestimated. Once an anomaly is detected, the possible consequent damages can be avoided or mitigated by carrying out necessary countermeasures.

### B. The TCP Reset Attack Detection

The TCP Reset Attack exploits an intrinsic weakness of the TCP protocol whose packets expose a header containing bits representing commands for the network interface. The TCP header, which is the initial part of an IP packet, contains a code of six bits, each of which is used for different purposes (e.g., SYN, FIN and RST are related to connection). TCP also manages to reorder the segments arriving at the recipient through a field of its header called sequence number [22]. As soon as a TCP RST packet with a coherent sequence number is received, the connection is broken. This RST packet is usually used when some errors are detected or in emergency situations.

The attacker observes the TCP packets of a specific connection between the two endpoints and then sends "forged" packets with the TCP reset flag enabled. The headers of the forged packet indicates, falsely, that it came from the genuine endpoint, not the forger (IP spoofing). In order to perform this kind of attack, the attacker must know the IP and the port where to send the forged packets. Properly formatted and forged, the TCP reset can be a very effective way to disrupt any TCP connection the forger can monitor.

Fig. 4 shows with a sequence diagram how the attack works: while the nodes A and B are communicating, the attacker periodically sends forged packets to node A. The forged packet, with a supposed sequence number and the RST bit enabled, is dropped if it does not have a sequence number such that the packet can be reordered together with the other packets received. However, if it has a sequence number

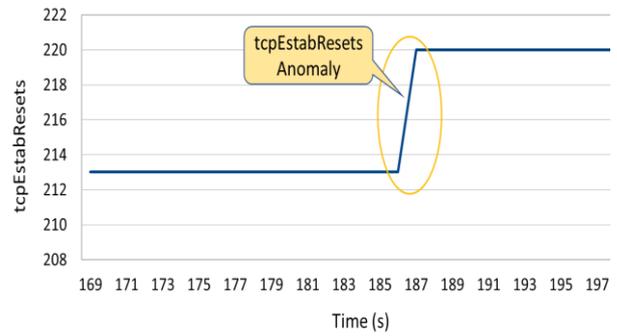

Fig. 3. The tcpEstabResets anomaly during the TCP Reset Attack

consistent with the last received packets, it is accepted causing the connection to be interrupted.

The TCP Reset Attack is basically detected by monitoring the *tcpEstabResets* measure; the SNMP agent increments this counter each time a connection is closed due to the reception of a TCP reset packet.

Since the attack can last for a relatively long time before the forged packet is accepted and the *tcpEstabResets* counter is updated, we monitored also an additional measure: *tcpRetransSegs*, representing the number of retransmissions executed by the ICT GW correlating it with a Socket Exception event. In this way, the attack is timely detected in both the following cases: i) a forged TCP RST packet deceives the ICT GW in few seconds, ii) the attack takes more time to trigger the reset but, due to the flooding effect, the connection is significantly slowed down.

In case of the RST packet deceives the receiver in few seconds, the *tcpEstabResets* value depicted in Fig. 3 grows by 7 units in about 1 second. Multiple resets happen because the ICT GW tries to reconnect to its destination but the attack persists and interrupts any reconnection until it causes a timeout that stops the communication service. The anomaly is correctly detected.

In case the attack takes more time to trigger the reset, since it does not break the connection yet and the *tcpEstabResets* is not affected, a successful detection of the attack is performed by monitoring the *tcpRetransSegs*, as shown in Fig. 5; the detection occurs in about 3 seconds, just after the prevention of communication and the occurrence of a Socket Exception.

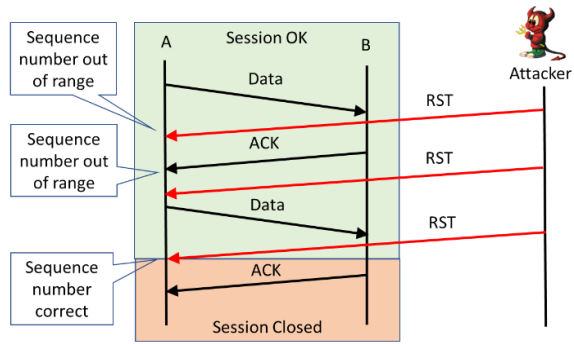

Fig. 4. The TCP Reset Attack sequence diagram

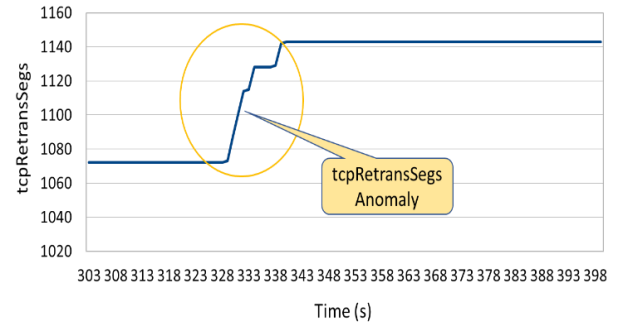

Fig. 5. The tcpRetransSegs anomaly during the TCP Reset Attack

| Name | EPL Code | Reasoning |
|---|---|---|
| Memory Overload | `select * from pattern [ every e=Event (e.getAdapterFlag()=false)] where (e.getSNMPvalue('hrStorageUsed')>62000)` | This pattern matches when the currently used RAM overcomes the specified threshold |
| CPU Overload | `select * from pattern [every e=Event (e.getAdapterFlag()=false)] where (e.getSNMPvalue ('hrProcessorLoad')>90)` | This pattern checks if a core processor workload overcomes the 90% threshold. |
| Tasks Number Overcoming | `select * from pattern [every e=Event(e.getMeasure()='hrSystemProcesses'] where (e.getSNMPvalue()>40)` | This pattern checks anomalies in the hrSystemProcesses number. When the number of processes overcomes 41 an anomaly is detected. |
| ICMP Flooding Attack | `select irstream * from pattern [every (e1=Event(e1.getMeasure()='icmpInEchos') -> e2=Event(e2.getMeasure()='icmpInEchos')) where timer:within(1 sec] where (e2.getSNMPvalue()-e1.getSNMPvalue()>3)` | When a flood is ongoing, the counter icmpInEchos grows up very quickly. If icmpInEchos grows over 3 ICMP messages in 1 second, the fault is detected. |
| TCP Reset Attack | `select * from pattern [every e1=Event (e1.getMeasure()='tcpEstabResets' -> e2= Event(e2.getMeasure()='tcpEstabResets') or e3=Event(e3.getMeasure()='tcpEstabResets' -> e4=Event(e4.getMeasure()='tcpEstabResets') and e5=Event (e5.getAdapterFlag()=false)] win:time(15 sec) where (e2.getSNMPvalue()-e1.getSNMPvalue()>0 and e4.getSNMPvalue()-e3.getSNMPvalue() >4 and e5.getSNMPvalue()='SocketException')` | This pattern recognizes a TCP reset attack when the counter tcpEstabResets grows up of at least 1 in 15 secs or tcpRetransSegs grows up at least of 5 in the same time window and a communication exception occurs. |

TABLE I. DETECTION PATTERNS AND EPL QUERIES DEFINED FOR THE EXPERIMENTS

In our experiments, during the normal working of the system, the growth of *tcpEstabReset* and *tcpRetransSegs* always remained below the thresholds and no false negatives have occurred.

### C. The ICMP Flood Attack Detection

By means of ICMP Flood Attack, also known as ping flood attack, the attacker attempts to overwhelm a targeted device with ICMP echo-request packets, causing the target to become inaccessible to normal traffic. The ICMP, is an Internet layer protocol used by network devices to communicate. The network diagnostic tools operate using ICMP. Commonly, ICMP echo-request and echo-reply messages are used to ping a network device for diagnosing its health or to test the connectivity between the sender and the device. During an attack, however, they are used to overload a target network with data packets. Executing a ping flood is dependent on attacker knowing the IP address of the target but it does not depend on knowing ports where services are deployed. Configuring the firewall to disallow pings can block attacks originating from outside of the network. However, blocking ping requests can have unintended consequences, including the inability to diagnose server problems.

The sequence diagram in Fig. 6 depicts the execution of an ICMP flooding on the node A while it is transmitting data to the node B. The attack produces ICMP requests with high frequency sent to the node A which tries to reply to the attacker emitting a series of ICMP replies. Due to the overload affecting the communication channel, data packets destinated to node B may not be emitted or may not reach their destination.

This attack is detected monitoring the SNMP counter named *icmpInEchos*, whose threshold is set to 3 ping/s after having experienced that this measure grows by around 100 pings in 30 seconds in case of a moderate flood of ping requests. Performing some similar experiments, we noticed that if the attack is active for a short time (i.e., 2 or 3 seconds) the communication between the data source and the ICT GW is not interrupted, and it is restored after the end of the attack. However, having set the threshold in this way, permits to detect also ICMP flooding attacks of short duration. If instead the attack lasts for more than 3 seconds, the connection between the two nodes will certainly be broken. Fig. 7 shows the anomaly during a strong ICMP flooding attack. In this case, the *icmpInEchoes* counter is incremented approximately of $0.4*10^5$ per second.

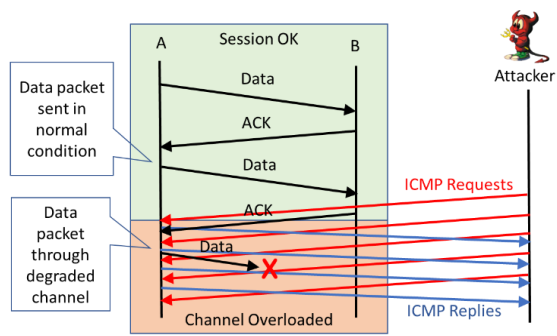

Fig. 6. The ICMP Flooding Attack

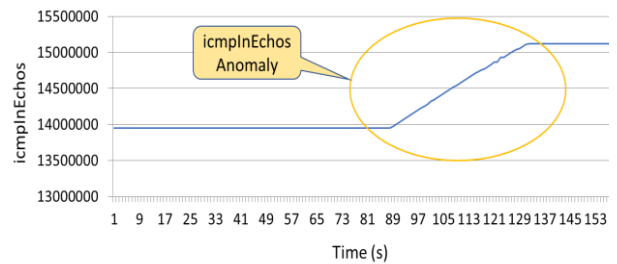

Fig. 7. The icmpInEchos anomaly during the ICMP flooding attack

*D. Detection Patterns*

Table 1 reports the detection pattern names used during the experiments, the related EPL code, and a short summary of the reasoning about the detection strategy.

## VI. CONCLUSIONS AND FUTURE WORKS

The fault and attack detection approach described in this paper combines the potentialities of SNMP with the CEP correlation capabilities. SNMP is able to provide online a huge set of different measurements with a secure communication protocol, while CEP allows to apply detection patterns with measurements correlation.

For brevity, we examined the most representative attacks affecting smart grids, the DoS attacks. Actually, the described approach allows to define countless attack detection patterns. New attack techniques can be thwarted by defining related new detection patters or attack evolutions can be addressed by modifying existing patterns. This feature makes it possible to organize an immediate reaction to attacks evolution in the application domain without making changes to the security infrastructure.

In addition, it is worth noting that the experiments described have considered only one node of the network consisting of a single networked machine. Although the approach, being extensible to many heterogeneous elements in the network, promises to be further enhanced for greater control of the network with greater capability in detecting faults. With reference to the current ICT technologies applied to smart grids, we are conducting a study aimed at understanding how the extension of this approach to the whole grid can interfere with communications performances. Since only few messages per minute are required to monitor a smart grid device and the SNMP messages relatively limited (e.g., 484 bytes) they do not interfere with the availability of the ICT infrastructure producing low impact on communication performances. However, this will be planned and experimented in further developments.

Finally, knowing the topology of the network, the anomalies correlation will also allow the location of accidental or malicious faults in the network. This is a study that will be deepened in the future and applied in the smart grid context.

ACKNOWLEDGMENT

This work has received funding from the European Union's Horizon 2020 research and innovation programme under grant agreement No 774145 for the Net2DG Project.